# Leveraging Data Mining Algorithms to Recommend Source Code Changes

AmirHossein Naghshzan[a,*], Saeed Khalilazar[a], Pierre Poilane[a], Olga Baysal[b], Latifa Guerrouj[a], Foutse Khomh[c]

[a]*Ecole de Technologie Supérieure, Montreal QC, Canada*
[b]*Carleton University, Ottawa ON, Canada*
[c]*Ecole Polytechnique de Montréal, Montreal QC, Canada*

**Abstract**

*Context:* Recent research has used data mining to develop techniques that can guide developers through source code changes. To the best of our knowledge, very few studies have investigated data mining techniques and–or compared their results with other algorithms or a baseline.

*Objectives:* This paper proposes an automatic method for recommending source code changes using four data mining algorithms. We not only use these algorithms to recommend source code changes, but we also conduct an empirical evaluation.

*Methods:* Our investigation includes seven open-source projects from which we extracted source change history at the file level. We used four widely data mining algorithms *i.e.*, Apriori, FP-Growth, Eclat, and Relim to compare the algorithms in terms of performance (Precision, Recall and F-measure) and execution time.

*Results:* Our findings provide empirical evidence that while some Frequent Pattern Mining algorithms, such as Apriori may outperform other algorithms in some cases, the results are not consistent throughout all the software projects, which is more likely due to the nature and characteristics of the studied projects, in particular their change history.

*Conclusion:* Apriori seems appropriate for large-scale projects, whereas

---
*Corresponding author.
Email addresses:* amirhossein.naghshzan.1@ens.etsmtl.ca (AmirHossein Naghshzan), a.baysal@carleton.ca (Olga Baysal), latifa.guerrouj@etsmtl.ca (Latifa Guerrouj), foutse.khomh@polymtl.ca (Foutse Khomh)

Eclat appears to be suitable for small-scale projects. Moreover, FP-Growth seems an efficient approach in terms of execution time.

## 1. Introduction

In recent years, developers have used recommendation systems to complete their daily tasks and activities in different phases of the development process. These tools prepare and provide developers with numerous artifacts that is appropriate and relevant to the context of their tasks such as third-party libraries, documentations and API function calls (Di Rocco et al., 2021). To avoid reinventing the wheel, recommendation systems could provide developers with useful information such as reusable code snippets, method invocations from external libraries, and the resolution of reported bugs. The ability of recommendation systems to search through large amounts of information improves their capability to present customized content to individual users.

The use of data mining algorithms for recommending source code changes has been steadily increasing in a variety of fields, including industrial, scientific, and commercial ones (Borg et al., 2017; Robillard et al., 2010; Ponzanelli et al., 2017). One of these fields in software engineering is system modifications. Software developers must consider modification tasks in terms of dependencies between different parts of the source code, such as files that were changed together regularly (Agrawal and Horgan, 1990; Parnas, 1972; Weiser, 1984). To help developers with their modifications tasks, we investigate several association rules mining algorithms to generate recommendations for source code changes using four data mining algorithms, namely Apriori, FP-Growth, Eclat and Relim. The primary goal of producing these file change recommendations is to assist developers during their software development and maintenance activities.

Our work is inspired by Zimmermann et al. (2005) and Ying et al. (2004) who used Apriori and FP-Growth algorithms to predict file change patterns. They proposed recommendation systems that suggest editing files based on mined software revision histories. The most edited files' association rules are mined, and then file-to-edit is recommended (Lee et al., 2015). Similar to these works, we predict source code file changes using data mining.

However, we investigate four data mining algorithms: Apriori, FP-Growth, Eclat and Relim with different configurations (support and confidence). Frequent Pattern Mining (FPM) algorithms can be divided into three categories: Join-Based, Tree-Based and Pattern Growth (Aggarwal et al., 2014). For this study we attempted to cover all three types by selecting well-known algorithms from different categories. Furthermore, we conducted an empirical study to evaluate the performance of those techniques on seven different projects. To the best of our knowledge, we are the first to investigate the use of four different data mining algorithms to predict changes in source code files. Our empirical study includes seven open-source projects: Eclipse, Elasticsearch, Rhino, SWT, Kotlin, Guava, and JabRef, from which we extracted the change history at the file level. We chose these projects because they are of varying sizes and scopes. Furthermore, they have a reasonable source code change history that we believe can be leveraged in such research.

The main research question that we address in our empirical investigation is as follows:

***RQ:*** *Are there any differences between the four studied data mining techniques, i.e., Apriori, FP-Growth, Eclat and Relim in terms of their performances when recommending source code file changes?*

To answer the above-mentioned research question, we quantified the performance of each examined data mining technique, for each project, in terms of Precision, Recall, and F-measure. Additionally, we perform pairwise comparisons between the different techniques for all projects in term of their performance when recommending source code file changes.

We derived six specific research questions from the above-mentioned research question. The hypotheses corresponding to our overarching research question will be discussed in detail in section 5.2.

- ***RQ1:*** *How does the Apriori perform compared to FP-Growth in terms of performance and execution time?*

- ***RQ2:*** *How does the Apriori perform compared to Eclat in terms of performance and execution time?*

- ***RQ3:*** *How does the Apriori perform compared to Relim in terms of performance and execution time?*

- **RQ4:** *How does the FP-Growth perform compared to Eclat in terms of performance and execution time?*

- **RQ5:** *How does the FP-Growth perform compared to Relim in terms of performance and execution time?*

- **RQ6:** *How does the Eclat perform compared to Relim in terms of performance and execution time?*

We empirically demonstrate that, the performance of Frequent Pattern Mining (FPM) algorithms highly depends on the characteristics of the studied projects like their change history. For instance, Apriori seems to perform well for large-scale projects within a reasonable time, while Eclat can be used for small-scale datasets. For large-scale projects not only Eclat's performance drops in terms of Precision and Recall, but the execution time also grows rapidly.

We believe that the research community working in this area can benefit from these findings when selecting the data mining algorithms to apply for source code changes or other engineering tasks.

## 2. Technical Background

In this section, we describe each data mining algorithm that we used in our study, *i.e.*, Apriori, FP-Growth, Eclat, and Relim. We selected these well-known algorithms from different categories to cover all the three types of Frequent Pattern Mining (FPM) algorithms, Join-Based, Tree-Based and Pattern Growth (Aggarwal et al., 2014).

*2.1. Apriori*

Agrawal et al. (1994) have proposed Apriori as the first data mining algorithm for frequent itemset mining. It is well-known as a reference and one of the most widely used association rule mining algorithms for quickly locating sets of items from transactions (Agrawal et al., 1994). In association rules, transactions are records of a dataset containing itemsets (Talia et al., 2016). To reduce the search space, this algorithm employs two steps: "join" and "prune". It is a level-by-level iterative approach to identifying the most frequent itemsets (Iqbal et al., 2020).

Apriori begins by generating candidates, or potential sets of frequently occurring items. In the second step, the list of candidates is used as input and checked to see if they are common. To accomplish this, each candidate from the transactions is validated by calculating its support, and if it is equal to or greater than the minimum support defined, the candidate is considered a frequent item and is added to the list of frequent sets. After that, an extended list of frequent candidates is generated, and the process is repeated (beginning with step 1) to build the next list of candidates, and so on. Finally, the algorithm terminates when the last generated list of candidates is empty.

*2.2. FP-Growth*

This algorithm is an improvement to the Apriori method proposed by Han et al. (2000). To solve the key drawback of algorithms such as Apriori, pattern-growth algorithms such as FP-Growth have made significant advances in the field (Fournier-Viger et al., 2017). Han et al. (2012) suggest that the algorithm employs a divide-and-conquer strategy. Without the need for candidate generation, a frequent pattern is generated. The FP-Growth algorithm represents the database as a tree known as a frequent pattern tree or FP tree.

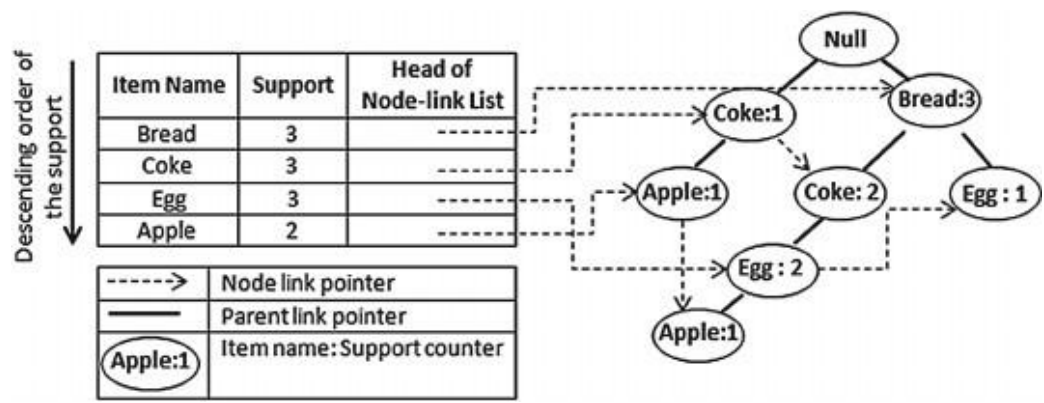

Figure 1: An FP-tree, adopted from Alsulim et al. (2015).

Rather than using Apriori's generate and test strategy, it builds an FP Tree. As Heaton (2016) mentioned Aprori is fundamentally a horizontal, breadth-first algorithm while FP-Growth's structure provides a vertical view of the data with a header table for each unique item that has support above the threshold support level. This header table comprises a linked list that traverses the tree to connect all nodes of the same kind. In addition to the vertical perspective, the header table provides FP-Growth with a horizontal view of the data. As illustrated in Figure 1, the FP tree is made up of a null root node and child nodes (Alsulim et al., 2015). The FP Growth algorithm focuses on fragmenting item paths and mining frequent patterns.

*2.3. Eclat*

Eclat, *i.e.*, Equivalence Class Transformation is a mining frequent itemsets algorithm introduced by Zaki et al. (1997). Eclat mines and groups all transactions with a specific itemset into a single record using a vertical data format. The database is scanned in the first step, and the data is converted from horizontal to vertical format. This process is illustrated in Table 1, Table 2, Table 3. It then produces frequent ($k + 1$)-itemsets by intersecting the transactions of the frequent $k$-itemsets. It will continue iterations until no more itemsets are generated.

Table 1: Transactional data in vertical format adopted from Chee et al. (2019).

| **Itemset** | **TID set** |
|:---:|:---:|
| I1 | T100, T400, T500, T700, T800, T900 |
| I2 | T100, T200, T300, T400, T600, T800, T900 |
| I3 | T300, T500, T600, T700, T800, T900 |
| I4 | T200, T400 |
| I5 | T100, T800 |

Unlike other algorithms, it is not necessary to scan the database multiple times. Eclat only scans the database once, and the data is transformed from horizontal to vertical (Chee et al., 2019). This depth-first search algorithm can easily calculate the support of an itemset (Pramod and Vyas, 2010) by using a vertical database layout. In comparison to the Apriori and FP-Growth algorithms, as the number of transactions increases the execution time of the Eclat algorithm increases, resulting eventually in a drop in performance (Yu and Wang, 2014).

*2.4. Relim*

Borgelt (2005) developed the Relim algorithm, which, like many other algorithms in the search for frequent elements, is based on the recursive deletion of elements from transactional database. Later, Wang et al. (2005) developed an extension of the algorithm to mine fuzzy frequent itemsets. Relim's main strength is its structure's simplicity, as everything is done in a recursive function (Fakir et al., 2020). As Borgelt (2005) stated, although the algorithm is very simple and does not require complex data structures, recursive elimination performs surprisingly well and it could be the method of choice if a straightforward and simple implementation is desired. On the other hand, Relim's structure is built in a time-consuming and resource-intensive manner. Relim algorithm differs from other extracting frequent items algorithms in terms of processing and data structure (Fakir et al., 2020).

Table 2: 2-Itemsets in vertical format adopted from Chee et al. (2019).

| Itemset | TID set |
| --- | --- |
| $I_1, I_2$ | T100, T400, T800, T900 |
| $I_1, I_3$ | T500, T700, T800, T900 |
| $I_1, I_4$ | T400 |
| $I_1, I_5$ | T100, T800 |
| $I_2, I_3$ | T300, T600, T800, T900 |
| $I_2, I_4$ | T200, T400 |
| $I_2, I_5$ | T100, T800 |
| $I_3, I_5$ | T800 |

Table 3: 3-Itemsets in vertical format adopted from Chee et al. (2019).

| Itemset | TID set |
| --- | --- |
| $I_1, I_2, I_3$ | T800, T900 |
| $I_1, I_2, I_5$ | T100, T800 |

Table 4 shows a brief summary of the main differences between the four investigated data mining algorithms.

Table 4: Pros and Cons of the investigated data mining algorithms.

| Algorithm | Advantages | Disadvantages |
|---|---|---|
| Apriori | - Applying an iterative level-wise search technique which results in (k+1)-itemsets discovery from k-itemsets (Chee et al., 2019). | - Being time-consuming while generating candidate.<br>- Needing numerous scans on the databases while performing.<br>- Generating redundant rules (Mythili and Shanavas, 2013). |
| Eclat | - The database is not required to be scanned multiple times in order to identify the (k+1)-itemsets (Chee et al., 2019).<br>- The database is also not required to be scanned multiple times to identify the support count of every frequent itemset (Chee et al., 2019). | - Takes extensive memory space and processing time for intersecting the itemsets (Chee et al., 2019). |
| FP-Growth | - Maintains the association information of all itemsets.<br>- Reducing target data to be scanned and searched (Chee et al., 2019). | - For large datasets, FP-tree construction is time-consuming (Chee et al., 2019). |
| Relim | - Simplicity of its structure (Applying recursive function) (Fakir et al., 2020). | - Consumes many system resources.<br>- Time-consuming (Fakir et al., 2020) |

## 3. Related Work

Several works in the literature on software engineering have used development history to build data mining-based recommenders. It has been used for the detection of bad smells in source code by Palomba et al. (2015). HIST (Historical Information for Smell Detection) is a system that uses data from change history to look for the presence of bad practices in the source. HIST's results have shown that using development history as a data source yields promising results.

Zimmermann et al. (2005) have proposed ROSE, which analyzes a project's entire history to predict the location of most likely future changes and employs the Apriori algorithm to compute association rules. They created an Eclipse plug-in that investigates source code changes retrieved from repositories to assist developers in identifying potential co-changes (Zimmermann et al., 2005). We increased the number of data mining algorithms in our study to four by including FP-Growth, Relim, and Eclat, whose performance is measured in terms of Precision, Recall, and F-measure. In addition, we empirically tested our approach with seven different open-source projects.

Malheiros et al. (2012) have introduced Mentor, which assists newcomers in resolving change requests by recommending source code files. It employs the Prediction by Partial Matching (PPM) algorithm, as well as some heuristics for data mining within a version control system and change request analysis to provide relevant source code recommendations that assist developers in dealing with change requests. It makes no difference whether the tool is written in C, C++, Java, or a combination of programming languages. It employs the Support Vector Machine (SVM) classifier to locate similar change requests (Malheiros et al., 2012).

Malik and Shakshuki (2010) have relied on different heuristics, including the analysis of related entities if they have similar files. By combining these heuristics, the researchers have demonstrated that source code changes can be used to generate recommendations for functions. In our study, recommendations are generated based on files changes rather than functions, using data mining techniques rather than heuristics. Using frequent itemsets mining techniques such as Apriori, FP-Growth, Eclat, and Relim, we can identify files changed together.

Other studies went a step further by incorporating the developer's work context. Ponzanelli et al. (2017), for example, used web browsing data navigated by developers, along with source code changes, and the semantic rela-tionships of the resources used by developers to help build a holistic recommendation system that can assist software developers. In our study, however, association rules are generated and reported based on changes made by developers using data mining algorithms.

Ying et al. (2004) have suggested a method for predicting file change patterns based on FP-Growth. Several other works used the change history to help developers make software changes. In essence, Uddin et al. (2011) has created a recommender that detects API changes by employing a temporary API usage pattern.

The above-mentioned studies did not take into account the temporal dimension. Robbes et al. (2010), on the other hand, conducted a study in which they recorded all developer interactions with the IDE, allowing them to track the time spent by each developer to perform their changes.

Nguyen et al. (2019) demonstrated FOCUS, a context-aware collaborative-filtering recommender system that mines open-source software (OSS) repositories to provide developers with API Function Calls and Usage Patterns. It employs the Expectation-Maximization (EM) algorithm in its structural form. Furthermore, Sisman and Kak (2012) focused on finding bugs in projects based on their development history. The authors focused specifically on the frequency of files that could be associated with defects and changes in history, which can then be used to estimate the prior probability that a given file would be the source of bugs.

We share with these research works the belief
that recommendation systems can help developers by guiding and supporting them as they perform software maintenance and evolution tasks. As we are interested to recommend files that changed together, our granularity is at the file level. Unlike previous work, we explore different data mining algorithms, *i.e.*, Apriori, FP-Growth, Eclat, and Relim and study their effect on recommending source code changes when using various support and confidence configurations. Additionally, we compared them in terms of performance (Precision, Recall, F-measure) and execution time and the results reported in section 6.

## 4. Methodology

This section describes the methodology used to address our research question. Our methodology consists of three major steps, as illustrated in Figure 2.

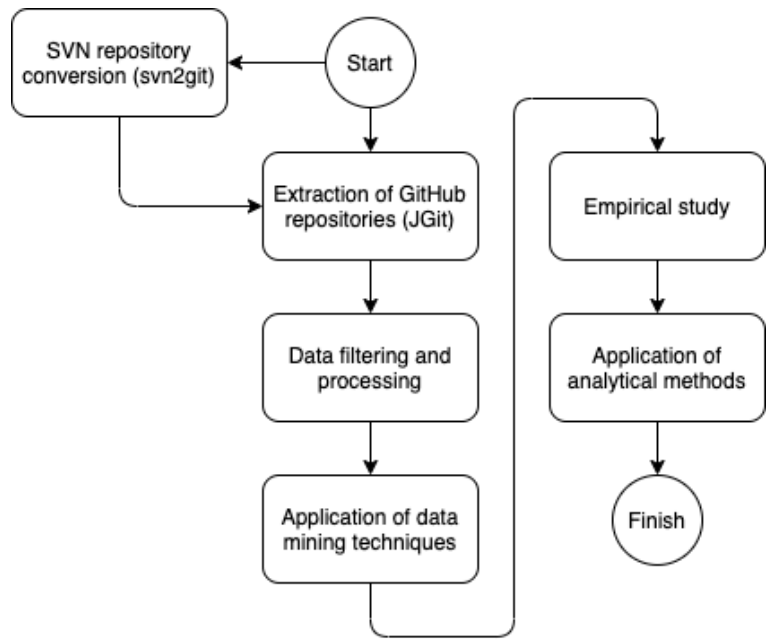

Figure 2: Overview of the followed methodology.

The first step consists of extracting the source code change for each studied project and saving it in a PostgreSQL database for data organization purposes. The second step entails filtering and generating training and test data. Finally, in the third step, the four data mining algorithms investigated have been applied, on the training data, to generate recommendations for source code change files. The recommendations generated are empirically validated against the test data. Following that, a comparison of the four algorithms, Apriori, FP-Growth, Eclat, and Relim, has been performed based on proper statistical tests.

*4.1. Extracting source code change history*

The first step in our methodology is to create a list of all Java source code file changes from a project repository using a change extraction program and the JGit tool[1] (Figure 2). Previously, if a project was only available in SVN, we had to use the svn2git tool[2] to migrate it to Git. This step begins with the creation of a local Git repository for the project under consideration. The history extraction program then goes through all the changes made to the repository in the past, and when the change involves one or more Java files (i.e., .Java files), the program selects the change. Various file-related information are then mined, including the change ID, change log, change date, type of modification performed (add, modify, rename, or delete), file name for each modified file, and a unique index for each file. Finally, using the various retrieved information, each selected file is placed into a PostGreSQL database. As a result, for each project, we form a database comprising all of the Java files that have been edited in the form of transactions. A transaction is a collection of files that have been changed in the same commit. Under Git, each file is identified by the same author, date, and message.

---

[1] https://www.eclipse.org/jgit/
[2] https://github.com/nirvdrum/svn2git

*4.2. Data filtering and processing*

The second step of our methodology entails filtering and processing the transactions retrieved from each project's PostgreSQL databases with a Perl script and the DBI module, as well as generating training and test datasets. For each project from which the local repository has been extracted in PostgreSQL, a Perl script will go through all the project transactions, start by filtering them, then by processing the transactions to organize them into transactions belonging to the training file and transactions belonging to the test file.

The first filter used only considers transactions that are linked to a single file. Transactions with only one file, whether for training or test data, are not useful for drawing any predictions because the data mining algorithms are used to extract frequent sets of files from the different files occurring in the same transactions. The second filter, ignores transactions with more than 100 files as used in previous studies (Ying et al., 2004; Hassan and Xie, 2012). Because the files contained within these transactions are not necessarily related to the files that have been modified to implement a new feature, these types of modifications and the information they give are irrelevant to the predictions we want.

Once the filters are applied, the processing phase selects the filtered transactions in order to split them into the training and test data. For dividing the data into training and test, we replicated a previous study (Zimmermann et al., 2005). In effect, for each studied change history of each project, we selected full months containing the last 1,000 transactions, but not more than 50% of all transactions as our test data (Zimmermann et al., 2005) and all the previous transactions considered as our training data. During this period, known as the evaluation period in the mining jargon, we tested each transaction to see if its files could be predicted from previous change history, *i.e.*, the training dataset.

For some recent projects, we have noticed that when selecting the last 1,000 transactions, the latter are spread over only a few months, while for older projects, the latter 1000 transactions may extend over several years. To overcome this challenge, we verified, for each project, if the last 1,000 transactions go beyond one year. If this is the case, the script only selects transactions from the last year. If, however, the transactions do not exceed one year, then the script rounds the date to the full month based on the earliest date in the thousand selected and most recent transactions as in (Zimmermann et al., 2005). As a result of this step, we create the training and test data for each examined project among our seven studied projects.

After filtering each project using the aforementioned procedure, the total number of transactions, the number of transactions after filtering, and the division of transactions into training and test transactions are shown in Table 5.

Table 5: Number of transactions for each project.

| **Project** | **Total transactions** | **After filtering** | **Training transactions** | **Test transactions** |
|---|---|---|---|---|
| Eclipse | 17,985 | 10,320 | 7,224 | 3,096 |
| Elasticsearch | 20,659 | 12,414 | 8,690 | 3,724 |
| Rhino | 2,840 | 1,406 | 984 | 422 |
| SWT | 21,588 | 8,260 | 5,782 | 2,478 |
| Kotlin | 25,710 | 15,115 | 10,580 | 4,535 |
| Guava | 3,887 | 2,399 | 1,679 | 720 |
| JabRef | 7,074 | 4,229 | 2,960 | 1,269 |

## 4.3. Recommending source code file changes

We used four different data mining algorithms, *i.e.*, Apriori, FP-Growth, Eclat, and Relim, on our training data set, which was previously produced in the second step of our process. To explore the relationships between source code file changes in our transactional databases and generate predictions of source code file changes. The file recommendations generated, from the training data set, have been validated against the test data.

The strength of an association rule is determined by two factors, *i.e.*, support and confidence. The frequency with which a specific rule appears in the database being mined is referred to as *support*. The number of times a given rule turns out to be true in practice is referred to as *confidence*. The minimum support and confidence parameters (Webb, 1989) must be considered when generating the change file recommendations, depending on the type of algorithm used. We empirically investigated various levels of support and confidence for each project. The reason is that our aim is to compare the performance of the four algorithms under consideration using various configurations.

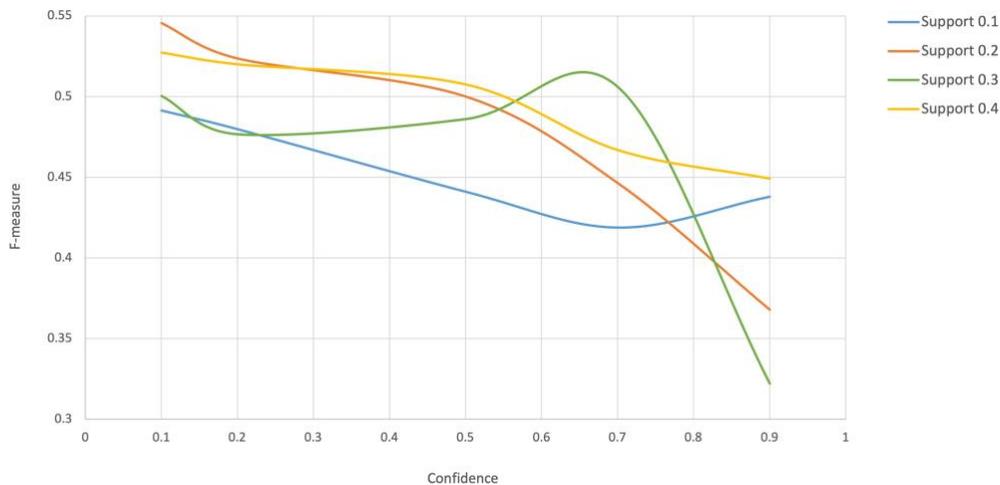

Figure 3: FP-Growth: varying support and confidence.

We have run several experiments and we have empirically investigated different pairs of support and confidence for the different considered data mining techniques when recommending source code changes. For example, Figure 3 shows the F-measure graph of the Eclipse project for the FP-Growth algorithm. Similar graphs have been generated for all examined instances of applied data mining techniques and investigated pairs of support and confidence. For space reasons, we present the rest of the graphs in our replication package. For each combination of minimum support and minimum confidence, the resulting F-measure is plotted. The x-axis represents the confidence while the y-axis represents the calculated F-measure and each line represents a specific value for minimum support. Subsequently, we have a line for each combination of minimum support and minimum confidence thresholds. As a result, we get four curves, one for each investigated support and confidence. As mentioned in a past research (Zimmermann et al., 2005), to choose the "best" support and confidence values for a certain project, graphs like these are required. As we can see in the figures, increasing confidence and support levels result in decreasing F-measure. Therefore we selected 0.1 confidence and 0.2 support as the optimum levels for our research. In the remainder of this paper, we have chosen values that are common across all projects, to facilitate comparison.

During our various experiments, we noticed that the lower the selected support and confidence parameters are, the greater the number of generated association rules, but at the expense of an increase in execution time. The same applies to the number of transactions as the execution time significantly increases when dealing with a large volume of data. Since the main purpose of producing these file change recommendations is to support and guide developers during their software development and maintenance activities, execution time must be normally as short as possible if real-time suggestions have to be made.

The output of the applied algorithms is a list of files that change together for each studied algorithm. Table 6 shows an example of generated rules that show files changed together for the Eclipse project.

Following the generation of source code change file recommendations for each project, our final task is to validate the relevance of the recommendations generated using each of our considered algorithms.

For each recommendation, we have compared if a correspondence exists in the test data of the concerned project. Statistics are generated from these various comparisons, and the performance of our algorithms is measured in terms of Precision, Recall, and F-measure. Other analytical methods, as presented Section 5, are then used to quantify and graphically represent these results, facilitating our discussion in Section 6 and Section 7.

Table 6: Example of generated rules for Eclipse.

| Rules | Changed Files |
| --- | --- |
| Rule 1 | org.eclipse.jdt.core/batch/compiler/batch/Main.java<br>org.eclipse.jdt.core/model/core/JavaCore.java<br>org.eclipse.jdt.core/compiler/compiler/ProblemReporter.java<br>org.eclipse.jdt.core/compiler/core/compiler/IProblem.java |
| Rule 2 | org.eclipse.jdt.core/compiler/compiler/ProblemReporter.java<br>org.eclipse.jdt.core/compiler/core/compiler/IProblem.java<br>org.eclipse.jdt.core/model/core/JavaCore.java |
| Rule 3 | org.eclipse.jdt.core/batch/compiler/batch/Main.java<br>org.eclipse.jdt.core/compiler/compiler/ProblemReporter.java<br>org.eclipse.jdt.core/compiler/core/compiler/IProblem.java |
| Rule 4 | org.eclipse.jdt.core/batch/compiler/batch/Main.java<br>org.eclipse.jdt.core/compiler/core/compiler/IProblem.java<br>org.eclipse.jdt.core/model/core/JavaCore.java |

## 5. Empirical Evaluation

To validate our research questions, like previous research (Naghshzan et al., 2021), we conducted an empirical study using the Basili framework, which consists of the following parts: definition, planning, operation, and interpretation (Basili et al., 1986).

*5.1. Definition and Planning of the Study*

The goal of our study is to investigate various data mining techniques algorithms on the recommendation of source code file changes.

The *Quality* focus is represented by the performance of the four investigated algorithms, *i.e.*, Apriori, FP-Growth, Eclat, and Relim, measured in terms of, Precision, Recall, and F-measure.

The *Perspective* is of researchers and practitioners who would like to build and–or use recommendation systems during their software engieering tasks.

The *Context* consists of seven open-source projects: Eclipse[3], Elasticsearch[4], Rhino[5], SWT[6], Kotlin[7], Guava[8], JabRef[9], and their source code history. The main characteristics of the studied projects are summarized in Table 7.

As it can be noticed, the evaluation period denotes the date of the first and last records in each project's databases. The *Records* column displays the total number of extracted records before transforming the commits data to transactions in the database corresponding to each examined project, while the *Contributors* column displays the total number of contributors

Table 7: Characteristics of investigated projects.

| **Project** | **Evaluation period** | **Records** | **Contributors** |
|---|---|---|---|
| Eclipse | 27/01/2001 – 24/04/2017 | 22,509 | 56 |
| Elasticsearch | 08/02/2010 – 01/06/2017 | 28,078 | 854 |
| Rhino | 21/04/1999 – 19/12/2016 | 3,383 | 31 |
| SWT | 02/08/2012 – 26/06/2017 | 25,585 | 53 |
| Kotlin | 10/12/2010 – 31/05/2017 | 35,115 | 197 |
| Guava | 15/09/2009 – 31/05/2017 | 5,654 | 189 |
| JabRef | 16/10/2003 – 25/04/2017 | 16,489 | 175 |

Also, Table 8 reports the investigated projects statistics including the number of releases, contributors, files, commits, and branches. Because the level of granularity in this study is a file, the most important measure to compare projects and the number of transactions will be the number of files.

---

[3] https://github.com/eclipse/eclipse.jdt.core
[4] https://github.com/elastic/elasticsearch
[5] https://github.com/mozilla/rhino
[6] https://www.eclipse.org/swt
[7] https://github.com/JetBrains/kotlin
[8] https://github.com/google/guava
[9] https://github.com/JabRef/jabref

Table 8: Projects statistics.

| Project | Releases | Contributors | Files | Commits | Branches |
| --- | --- | --- | --- | --- | --- |
| Eclipse | 87 | 325 | 8.6k | 1230 | 129 |
| Elasticsearch | 74 | 1712 | 22.5k | 62,868 | 163 |
| Rhino | 15 | 76 | 4.6k | 4090 | 51 |
| SWT | 2 | 12 | 1k | 466 | 10 |
| Kotlin | 179 | 532 | 65.6k | 91,069 | 4k |
| Guava | 34 | 271 | 3k | 5,735 | 19 |
| JabRef | 33 | 403 | 13k | 16,888 | 20 |

Elasticsearch and Kotlin are the largest projects in terms of files and contributors, according to this table.

*5.2. Research Questions*

By conducting this empirical evaluation, we are investigating 6 questions derived from our main research question. The questions and the null and alternative hypotheses corresponding to the research question are formulated as follows:

1. **RQ1:** *How does the Apriori perform compared to FP-Growth in terms of performance and execution time?*
   - $H_{o\_1}$: There is no statistically significant difference between Apriori and FP-Growth in terms of their performance and execution time.
   - $H_{a\_1}$: There is statistically a significant difference between Apriori and FP-Growth in terms of their performance and execution time.

2. **RQ2:** *How does the Apriori perform compared to Eclat in terms of performance and execution time?*
   - $H_{o\_2}$: There is no statistically significant difference between Apriori and Eclat in terms of their performance and execution time.

- $H_{a\_2}$: There is statistically a significant difference between Apriori and Eclat in terms of their performance and execution time.

3. **RQ3:** *How does the Apriori perform compared to Relim in terms of performance and execution time?*

    - $H_{o\_3}$: There is no statistically significant difference between Apriori and Relim in terms of their performance and execution time.
    - $H_{a\_3}$: There is statistically a significant difference between Apriori and Relim in terms of their performance and execution time.

4. **RQ4:** *How does the FP-Growth perform compared to Eclat in terms of performance and execution time?*

    - $H_{o\_4}$: There is no statistically significant difference between FP-Growth and Eclat in terms of their performance and execution time.
    - $H_{a\_4}$: There is statistically a significant difference between FP-Growth and Eclat in terms of their performance and execution time.

5. **RQ5:** *How does the FP-Growth perform compared to Relim in terms of performance and execution time?*

    - $H_{o\_5}$: There is no statistically significant difference between FP-Growth and Relim in terms of their performance and execution time.
    - $H_{a\_5}$: There is statistically a significant difference between FP-Growth and Relim in terms of their performance and execution time.

6. **RQ6:** *How does the Eclat perform compared to Relim in terms of performance and execution time?*

    - $H_{o\_6}$: There is no statistically significant difference between Eclat and Relim in terms of their performance and execution time.
    - $H_{a\_6}$: There is statistically a significant difference between Eclat and Relim in terms of their performance and execution time.

## 5.3. Variables selection

In this section, we present the variables related to our empirical investigation as well as the metrics used to assess the performance of the algorithms under consideration.

The type of data mining algorithm used is the main independent variable in our study. This factor has four different treatment, *i.e.*, Apriori, FP-Growth, Eclat, and Relim.

The dependent variable in our study is the performance of the data mining algorithms measured in terms of Precision, Recall, and F-measure as well as execution time.

We compute the peformance in terms of precision, recall, and F-measure using the method described in previous research (Zimmermann et al., 2005; Ying et al., 2004). In effect, for each transaction of the test data set, we check whether its entities can be predicted from earlier history.

We create a test case ($t$) that consists of a query ($q$) and an expected outcome ($B$). A query is simply a file from a transaction in the test data (Zimmermann et al., 2005) and the an expected outcome ($B_t$) includes all items and files in those transactions but not the query itself. Let us consider that ($A_t$) represents the rules/recommendations in our case generated for the query ($q$) by one of our data mining algorithms (Apriori, FP-Growth, Eclat, or Relim). We assess the test case ($q$) using the Precision ($P_t$), Recall ($R_t$), and F-measure ($F_t$). F-measure is the harmonic mean of Precision and Recall (van Rijsbergen, 1979):

$$P_t = \frac{|A_t \cap B_t|}{|A_t|}, \; R_t = \frac{|A_t \cap B_t|}{|B_t|}, \; F_t = \frac{2.P_t.R_t}{P_t+R_t}$$

To measure the overall performance of an algorithm, we computed the mean value of the Precision, Recall, and F-measure of queries with the selected min confidence and min support.

Furthermore, to estimate the execution time, we use Python *Timer* package to calculate the run-time of each algorithm. We start the timer right before executing the data mining algorithms and stop it as soon as the rules are generated. Clearly, the execution time is the time taken by each applied algorithm to generate the rules, *i.e.*, recommendations of source code change files for each studied project.

## 5.4. Analysis method

Our method of analysis relies on both descriptive statistics and statistical analyses. We used boxplots to present the results for each descriptive statistic (McGill et al., 1978).

We used the Shapiro-Wilk normality test (Shapiro and Wilk, 1965) to determine whether or not our data follows a normal distribution. When the p-value is greater than 0.05, it indicates that the data distribution is not significantly different from the normal distribution.

Table 9: Results of the Shapiro-Wilk normality test.

| Project | Precision | Recall | F-measure |
| --- | --- | --- | --- |
| Eclipse | <0.0001 | 0.1103 | 0.02614 |
| Elasticsearch | <0.0001 | 0.0009518 | <0.0001 |
| Guava | <0.0001 | <0.0001 | <0.0001 |
| JabRef | <0.0001 | <0.0001 | <0.0001 |
| Kotlin | 0.00708 | 0.01017 | 0.001215 |
| Rhino | 0.007473 | 0.001503 | 0.1342 |
| SWT | 0.0001218 | <0.0001 | 0.003301 |

As shown in Table 9, almost all of the performance measures of the various projects that we examined do not follow a normal distribution. As a result, we must use non-parametric statistical tests on our data.

We compared the Precision, Recall, and F-measure using the Wilcoxon paired test (Wilcoxon, 1947), which is a non-parametric test for pair-wise median comparison. The Wilcoxon test determines whether or not the median difference between two data mining algorithms is zero. Because the Wilcoxon test was used several times, p-values must be adjusted. The Holm correction method (Holm, 1979) was used to achieve this goal. This procedure sorts the p-value results from n tests in ascending order of magnitude, multiplying the smallest by $n$, the next by $n - 1$, and so on. The results are interpreted as statistically significant at $\alpha = 5\%$. The non-parametric

Cliff's delta is then calculated to determine the magnitude of the effect between two different data mining techniques. Cliff's delta ranges from -1 to 1, and its values represent various levels of effectiveness, with $0.474 \leq |d|$ is large, $0.33 \leq |d| < 0.474$ considered medium, $0.147 \leq |d| < 0.33$ is small, and $|d| < 0.147$ is insignificant.

## 6. Results

In this section, we present the descriptive statistics of the investigated data mining algorithms along with the boxplots corresponding to the performance of the studied data mining algorithms, *i.e.*, Apriori, FP-Growth, Eclat, and Relim as well as their execution time for generating rules of each studied project.

In the following, for each project, first, we will present the overall performance of algorithms and compare them with each other in terms of Precision, Recall, and F-measure. Additionally, we present all the adjusted p-values and Cliff's Delta values for performance measures. Then we will present the average processing time of each algorithm for that specific project and provide an overall conclusion at the end.

*6.1. Are there any differences between the four data mining techniques considered, i.e., Apriori, FP-Growth, Eclat and Relim in terms of their performance and execution time when recommending source code file changes for Eclipse?*

Figure 4 shows that Apriori outperforms the other three data mining techniques in terms of Precision, while Eclat, FP-Growth, and Relim are nearly identical and only slightly different.

In terms of Recall, the boxplots in Figure 5 show that Apriori outperforms other algorithms. Furthermore, it can be seen that Eclat outperforms Relim and FP-Growth, while there is no discernible difference between Relim and FP-Growth.

Finally, Figure 6 demonstrates that Apriori outperforms the other three data mining techniques in terms of F-measure *i.e.*, Eclat, FP-Growth, and Relim. Furthermore, while Eclat, FP-Growth, and Relim yield the same trend for precision, *i.e.*, they are only slightly different from each other in terms of F-measure.

While boxplots remain a simple way to illustrate the results visually, we need rigorous statistical tests to investigate if there are any significant differences between algorithms. In the following, we present the result of this test for the Eclipse project.

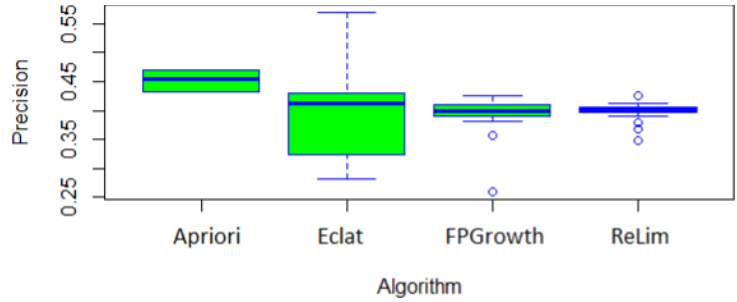

Figure 4: Boxplots of Precision for the Eclipse project

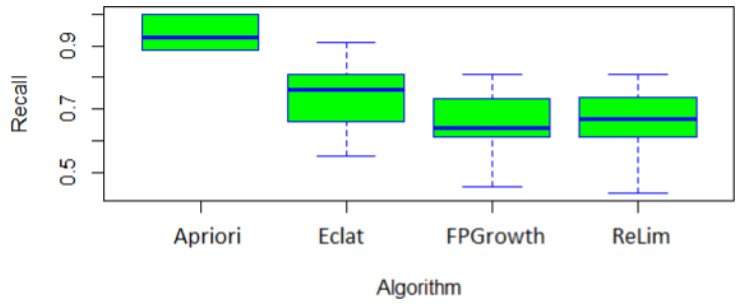

Figure 5: Boxplots of Recall for the Eclipse project

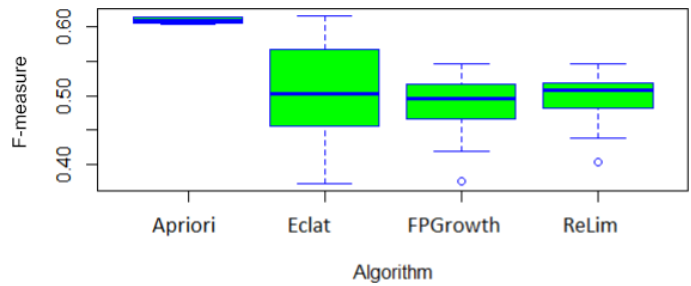

Figure 6: Boxplots of F-measure for the Eclipse project

Table 10 summarizes the p-values and Cliff's Delta values for each pair of comparisons in the Eclipse project. As it can be noticed for precision, there are statistically significant differences between Apriori and FP-Growth (p-value = 0.031 and Cliff's delta is large) as well as statistically significant differences between Relim and Apriori (p-value = 0.031 and Cliff's delta is large).

Table 10: Adjusted P-values and Cliff's Delta for paired comparison between the four studied data mining algorithms for Eclipse.

| Algorithm | Precision | | Recall | | F-measure | |
|---|---|---|---|---|---|---|
| | P-value | Cliff's Delta | P-value | Cliff's Delta | P-value | Cliff's Delta |
| Apriori - FP-Growth | 0.031 | **1 (large)** | 0.031 | **1 (large)** | 0.031 | **1 (large)** |
| Relim - FP-Growth | 0.62 | 0.04 (negligible) | 0.95 | 0.059 (negligible) | 0.95 | 0.12 (negligible) |
| Relim - Apriori | 0.031 | **-1 (large)** | 0.031 | **-1 (large)** | 0.031 | **-1 (large)** |
| Apriori - Eclat | 0.43 | **0.79 (large)** | 0.031 | **0.95 (large)** | 0.06 | **0.89 (large)** |
| Relim - Eclat | 0.74 | -0.2 (small) | 0.00077 | -0.38 (medium) | 0.52 | -0.15 (small) |
| FP-Growth - Eclat | 0.92 | -0.23 (small) | 0.00074 | -0.41 (medium) | 0.14 | -0.18 (small) |

Regarding recall, there are statistically significant differences between Apriori and FP-Growth as well as Relim and Apriori (p-value =0.03125, and Cliff's delta is large). Table 10 also shows statistically significant differences between Apriori and Eclat (p-value = 0.031 and Cliff's delta is large) and Relim and Eclat (p-value =0.0007766 and Cliff's delta is medium). Furthermore, the differences between FP-Growth and Eclat are statistically significant (p-value =0.0007482, and Cliff's delta is medium).

In terms of F-measure, as it can be seen in Table 10, there are statistically significant differences between Apriori and FP-Growth as well as Relim and Apriori (p-value = 0.031 and Cliff's delta is large).

Table 11: Average execution time for Eclipse.

| **Algorithm** | **Time (seconds)** |
| --- | --- |
| Apriori | 4,120 |
| FP-Growth | 2,550 |
| Relim | 8,410 |
| Eclat | 10,345 |

Our comparison involves also the execution time that each algorithm took to generate rules, *i.e.*, sources code change file recommendations. In effect, as shown in Table 11, FP-Growth with 2,550 seconds is the fastest algorithm, while Eclat with a runtime of 10,345 is the slowest one. For Apriori and Relim, it took 4,120 and 8,410 respectively to generate source code change rules.

Overall, we can conclude that there are statistically significant differences in Precision, Recall, and F-measure with a large effect-size measure between Apriori and FP-Growth as well as Apriori and Relim. Furthermore, when recommending source code change files, Eclat outperforms Relim and FP-Growth in terms of Recall with a medium effect-size measure. Moreover, in terms of execution time, FP-Growth is the most efficient algorithm while Apriori takes about 2x time as the second algorithm and Eclat seems to be the slowest one. Therefore, We reject the null hypotheses H0-1 and H0-3 for the comparison between Apriori and FP-Growth as well as Relim and Apriori.

*6.2.* Are there any differences between the four data mining techniques considered, *i.e.*, Apriori, FP-Growth, Eclat and Relim in terms of their performance and exceution time when recommending source code file changes for Elasticsearch?

Figure 7 shows that Apriori outperforms the other three data mining techniques in terms of Precision. Furthermore, it can be seen that Eclat and Relim outperform FP-Growth.

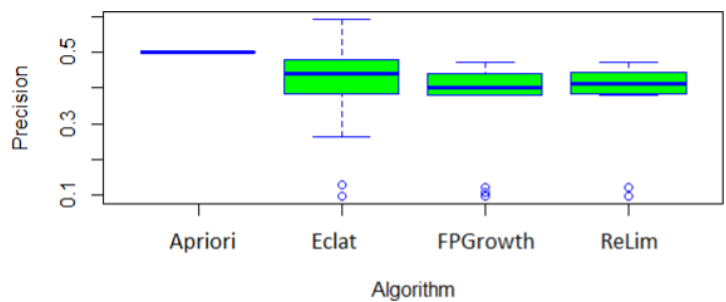

Figure 7: Boxplots of Precision for the Elasticsearch project

In terms of Recall, the boxplots in Figure 8 show that Apriori outperforms the other three data mining techniques in terms of Recall. Furthermore, there is no discernible difference between the performance of Eclat, FP-Growth, and Relim.

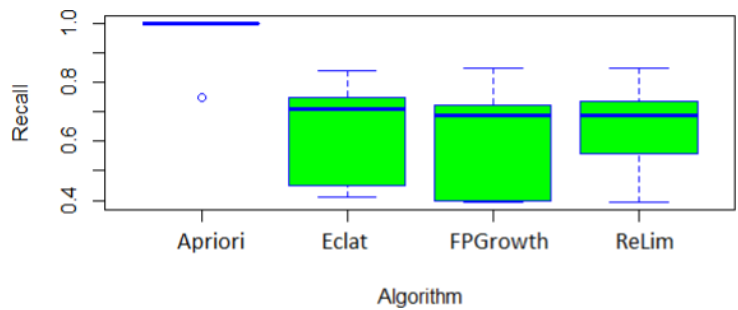

Figure 8: Boxplots of Recall for the Elasticsearch project.

Finally, Figure 9 demonstrates that Apriori outperforms the other three data mining techniques in terms of F-measure. Furthermore, it can be seen that Eclat and Relim outperform FP-Growth.

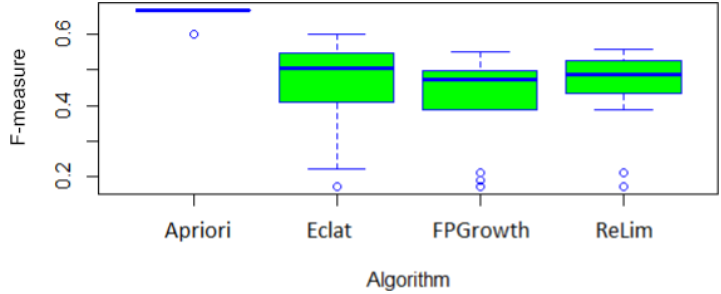

Figure 9: Boxplots of F-measure for the Elasticsearch project.

As shown in Table 12, there are statistically significant differences between Apriori and FP-Growth (p-value =0.031 and Cliff's delta is large) and Relim and FP-Growth (p-value =0.037 and Cliff's delta is small) for the Elasticsearch project. There are also statistically significant differences between Relim and Apriori (p-value =0.031 and Cliff's delta is large) and Apriori and Eclat (p-value =0.031 and Cliff's delta is large). Furthermore, we found statistically significant differences between FP-Growth and Eclat (p-value =0.009, with a small Cliff's delta).

In terms of Recall, there are statistically significant differences between Apriori and FP-Growth, as well as between Relim and Apriori (p-value =0.031, and Cliff's delta is large). Furthermore, there are statistically significant differences between Apriori and Eclat (p-value =0.031, with a large Cliff's delta) and FP-Growth and Eclat (p-value =0.0076, with a small Cliff's delta).

In terms of the F-measure, Table 12 shows statistically significant differences between Apriori and FP-Growth, as well as Relim and Apriori (p-value =0.031 and Cliff's delta is large) and Relim and FP-Growth (p-value =0.03 and Cliff's delta is small). Table 1-10 also reveals statistically significant differences between Apriori and Eclat (p-value =0.031 and Cliff's delta is large) and FP-Growth and Eclat (p-value =0. 00058 and Cliff's delta is small).

As shown in Table 13 FP-Growth with a runtime of 4,280 is the the fastest algorithm, while Eclat with 56,384 seconds is the slowest one. Also, for Apriori and Relim, it took 5,163 and 36,290 respectively to generate the rules.

For Elasticsearch, we conclude that there are statistically significant dif-

Table 12: Adjusted P-values and Cliff's Delta for paired comparison between the four studied data mining algorithms for Elasticsearch.

|  | Precision | | Recall | | F-measure | |
|---|---|---|---|---|---|---|
| Algorithm | P-value | Cliff's Delta | P-value | Cliff's Delta | P-value | Cliff's Delta |
| Apriori - FP-Growth | 0.031 | **1 (large)** | 0.031 | **0.95 (large)** | 0.031 | **1 (large)** |
| Relim - FP-Growth | 0.037 | 0.17 (small) | 0.1 | 0.083 (negligible) | 0.03 | 0.18 (small) |
| Relim - Apriori | 0.031 | **-1 (large)** | 0.031 | **-0.96 (large)** | 0.031 | **-1 (large)** |
| Apriori - Eclat | 0.031 | **0.89 (large)** | 0.031 | **0.93 (large)** | 0.031 | **1 (large)** |
| Relim - Eclat | 0.21 | -0.12 (negligible) | 0.27 | -0.12 (negligible) | 0.06 | -0.12 (negligible) |
| FP-Growth - Eclat | 0.009 | -0.3 (small) | 0.0076 | -0.18 (small) | 0.00058 | -0.26 (small) |

ferences for almost all instances of comparisons between algorithms (for the triple Precision, Recall, and F-measure) with effect sizes ranging from small to large, except Relim and Eclat, for which no statistically significant differences have been obtained between these two data mining algorithms. Also, in terms of execution time, as described in Table 13 while Apriori's execution time is close to FP-Growth, Relim and Eclat's execution times are 9 times and 14 times more than FP-Growth. We, therefore, reject the null hypothesis, for this project, for all instances of comparisons except for Relim

Table 13: Average execution time for Elasticsearch.

| Algorithm | Time (seconds) |
|---|---|
| Apriori | 5,163 |
| FP-Growth | 4,280 |
| Relim | 36,290 |
| Eclat | 56,384 |

– Eclat.

6.3. Are there any differences between the four data mining techniques considered, i.e., Apriori, FP-Growth, Eclat and Relim in terms of their performance and execution time when recommending source code file changes for Guava?

Figure 10 shows that there are no significant differences in Precision between the four data mining techniques.

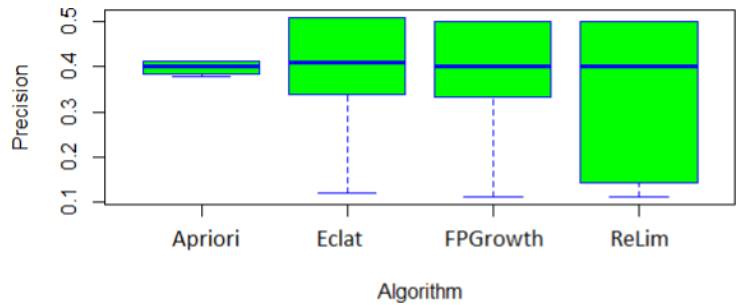

Figure 10: Boxplots of Precision for the Guava project.

In terms of Recall, the boxplots in Figure 11 show that Eclat and FP-Growth outperform Relim and Apriori.

Furthermore, the boxplots in Figure 12 show that there are no significant differences in F-measure performance between the four data mining techniques.

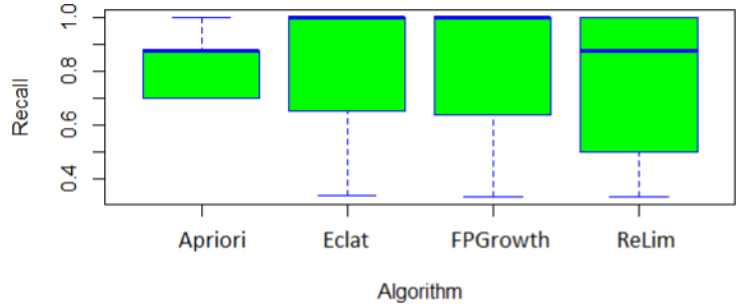

Figure 11: Boxplots of Recall for the Guava project.

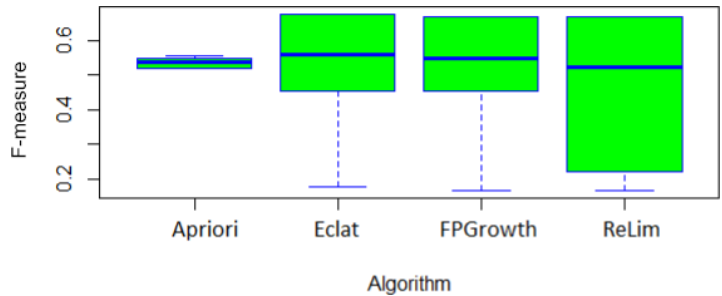

Figure 12: Boxplots of F-measure for the Guava project.

For the Guava project, as shown in Table 14, there are statistically significant differences between FP-Growth and Eclat for the Guava project (p-value =0.00027 and Cliff's delta is small), as well as statistically significant differences between Relim and Eclat (p-value =0.00027 and Cliff's delta is small).

For what concerns the recall,, Table 14 shows statistically significant differences between Apriori and FP-Growth, as well as Relim and Apriori (p-value =0.035 with a negligible Cliff's delta). Furthermore, with a negligible Cliff's delta, there are statistically significant differences between Apriori and Eclat, Relim – Eclat, and FP-Growth – Eclat.

In terms of F-measure, Table 14 shows statistically significant differences between Apriori and FP-Growth (p-value =0.035), Relim and Apriori (p-value =0.035), and Apriori and Eclat (p-value =0.058), but with a negligible Cliff's delta. Furthermore, statistically significant differences exist between Relim and Eclat (p-value = 0.00018) and FP-Growth − Eclat (p-value = 0.00018).

Table 14: Adjusted P-values and Cliff's Delta for paired comparison between the four studied data mining algorithms for Guava.

| Algorithm | Precision | | Recall | | F-measure | |
| --- | --- | --- | --- | --- | --- | --- |
| | P-value | Cliff's Delta | P-value | Cliff's Delta | P-value | Cliff's Delta |
| Apriori - FP-Growth | 0.29 | -0.09 (negligible) | 0.035 | -0.13 (negligible) | 0.035 | -0.07 (negligible) |
| Relim - FP-Growth | 0.37 | -0.056 (negligible) | 1 | -0.07 (negligible) | 0.37 | -0.06 (negligible) |
| Relim - Apriori | 0.29 | 0.037 (negligible) | 0.035 | 0.028 (negligible) | 0.035 | -0.037 (negligible) |
| Apriori - Eclat | 0.28 | -0.17 (small) | 0.035 | -0.13 (negligible) | 0.058 | -0.11 (negligible) |
| Relim - Eclat | 0.00027 | -0.31 (small) | 0.009 | -0.13 (negligible) | 0.00018 | -0.31 (small) |
| FP-Growth - Eclat | 0.00027 | -0.26 (small) | 0.01 | -0.05 (negligible) | 0.00018 | -0.25 (small) |

Considering execution time, while the Relim with 930 seconds is the slowest algorithm (Table 15), Apriori and FP-Growth have close results. The execution time for Apriori and FP-Growth is 140 and 230 seconds. While it took 783 seconds for Eclat to generate the rules.

Table 15: Average execution time for Guava.

| Algorithm | Time (seconds) |
|---|---|
| Apriori | 140 |
| FP-Growth | 230 |
| Relim | 930 |
| Eclat | 783 |

Overall, for Guava, we can conclude that there are statistically significant differences in Precision, Recall, and F-measure between Relim and Eclat, as well as FP-Growth and Eclat, but with a negligible to small effect-size measure. Other comparisons show statistically significant differences for the Recall and F-measure components, but the effect-size measure is negligible. Likewise, speaking about execution time all algorithms have close results especially FP-Growth and Apriori.

6.4. Are there any differences between the four data mining techniques considered, i.e., Apriori, FP-Growth, Eclat and Relim in terms of their performance and execution when recommending source code file changes for JabRef?

In terms of Precision, Figure 13 shows that Apriori outperforms other data mining techniques.

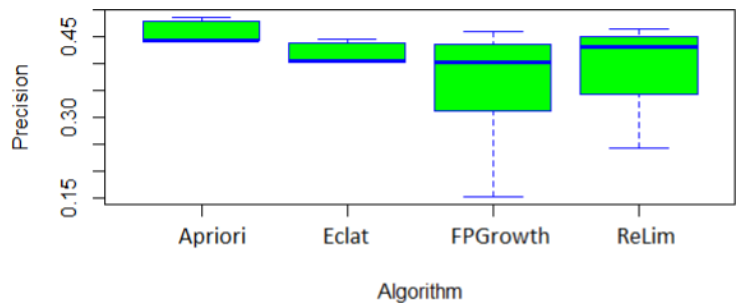

Figure 13: Boxplots of Precision for the JabRef project.

In addition, the boxplots in Figure 14 show that Apriori outperforms other data mining techniques in terms of Recall.

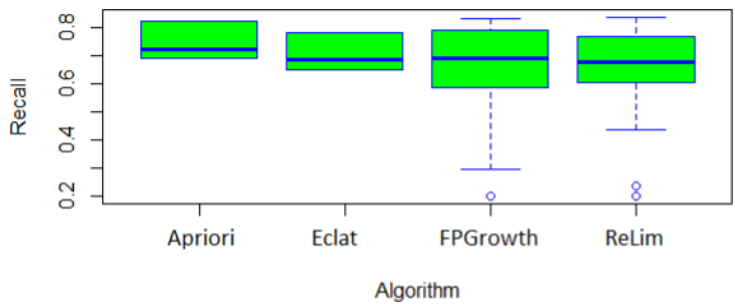

Figure 14: Boxplots of Recall for the JabRef project.

Finally, Figure 15 shows that Apriori outperforms other data mining techniques in terms of F-measure.

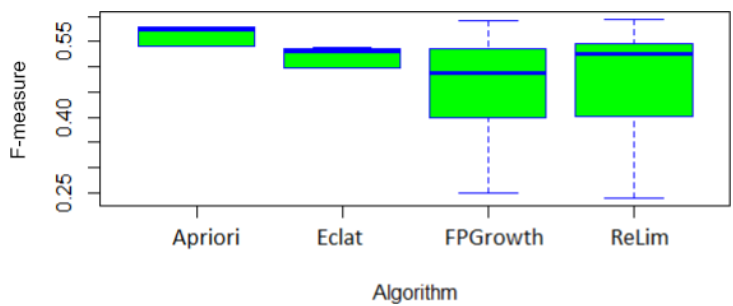

Figure 15: Boxplots of F-measure for the JabRef project.

As shown in Table 16, there are statistically significant differences between Apriori and Eclat for the JabRef project (p-value =0.035 and Cliff's delta is large).

In terms of Recall, Table 16 shows that there are statistically significant differences between Relim and Apriori (p-value =0.031 and Cliff's delta is small) as well as statistically significant differences between Apriori and Eclat (p-value =0.035 and Cliff's delta is medium). There are also differences between FP-Growth and Eclat, as well as Relim and Eclat (p-value =0.031, while Cliff's delta is negligible).

In terms of the F-measure, as shown in Table 16, there are statistically significant differences between Apriori and Eclat (p-value =0.035 and a large Cliff's delta) as well as Relim and Apriori (p-value =0.031 and a large Cliff's delta).

Table 16: Adjusted P-values and Cliff's Delta for paired comparison between the four studied data mining algorithms for JabRef.

|  | Precision | | Recall | | F-measure | |
| --- | --- | --- | --- | --- | --- | --- |
| Algorithm | P-value | Cliff's Delta | P-value | Cliff's Delta | P-value | Cliff's Delta |
| Apriori - FP-Growth | 0.16 | **0.7 (large)** | 0.29 | 0.31 (small) | 0.29 | **0.74 (large)** |
| Relim - FP-Growth | 0.15 | 0.2 (small) | 0.13 | 0.09 (negligible) | 0.127 | 0.16 (small) |
| Relim - Apriori | 0.15 | **-0.48 (large)** | 0.031 | 0.28 (small) | 0.031 | **-0.59 (large)** |
| Apriori - Eclat | 0.035 | **0.78 (large)** | 0.035 | 0.44 (medium) | 0.035 | **1 (large)**2 |
| Relim - Eclat | 0.84 | 0.17 (small) | 0.031 | 0.06 (negligible) | 0.31 | 0 (negligible) |
| FP-Growth - Eclat | 0.84 | -0.18 (small) | 0.031 | -0.06 (negligible) | 0.22 | -0.08 (negligible) |

As shown in Table 17, FP-Growth yields a runtime of 253 seconds making the the fastest algorithm among those investigated, while Eclat with 4,427 seconds is the slowest one. Apriori and Relim took 890 and 3,945 respectively to make source code changes recommendations.

Table 17: Average execution time for JabRef.

| Algorithm | Time (seconds) |
|---|---|
| Apriori | 890 |
| FP-Growth | 253 |
| Relim | 3,945 |
| Eclat | 4,427 |

Overall, for JabRef, we reject the null hypothesis H0-14 for the comparison of Apriori and Eclat because there are statistically significant differences in Precision, Recall, and F-measure with a medium to large effect size. Other comparisons, such as the one between Relim and Apriori, produce statistically significant differences, but only for Recall and F-measure. In terms of execution time, while FP-Growth and Apriori's results are close, and Eclat are about 20x slower than FP-Growth but close to Relim's execution time.

6.5. Are there any differences between the four data mining techniques considered, i.e., Apriori, FP-Growth, Eclat and Relim in terms of their performance and execution time when recommending source code file changes for Kotlin?

Figure 16 shows that Apriori outperforms other data mining techniques in terms of Precision.

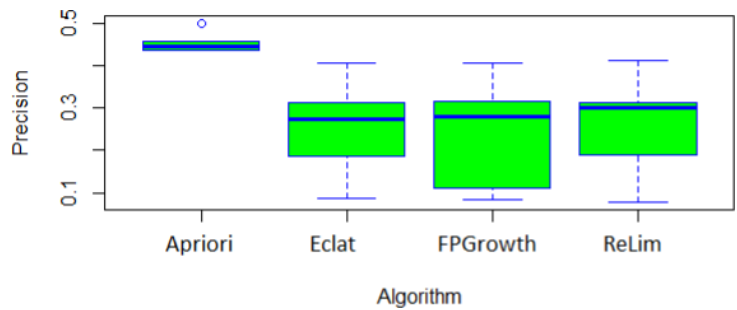

Figure 16: Boxplots of Precision for the Kotlin project.

And the boxplots in Figure 17 show that Apriori outperforms other data mining techniques in terms of Recall.

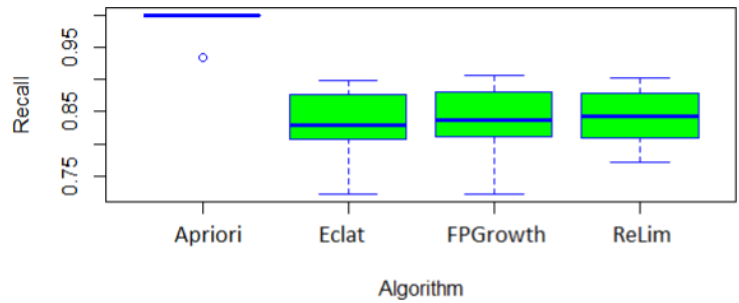

Figure 17: Boxplots of Recall for the Kotlin project.

Furthermore, Figure 18 shows that Apriori outperforms other data mining techniques in terms of F-measure.

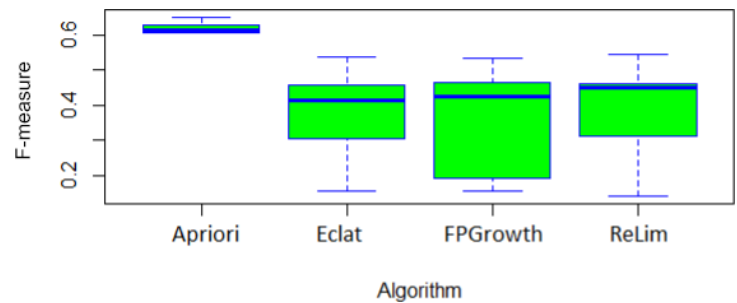

Figure 18: Boxplots of F-measure for the Kotlin project.

As shown in Table 18, there are statistically significant differences between Apriori and FP-Growth (p-value =0.03125 and Cliff's delta is large), Relim and Apriori (p-value =0.031 and Cliff's delta is large), and Apriori and Eclat (p-value =0.031 and Cliff's delta is large) for the Kotlin project. For these comparisons, the same trends are observed for Recall and F-measure.

For Kotlin, based on Table 19, Eclat with 27,434 seconds of run-time is the slowest algorithm compared to other algorithms. However, FP-Growth has the best execution time with 1,030 seconds and for Apriori and Relim it took 3,550 and 15,1918 seconds to generate the rules.

Table 18: Adjusted P-values and Cliff's Delta for paired comparison between the four studied data mining algorithms for Kotlin.

|  | Precision | | Recall | | F-measure | |
| --- | --- | --- | --- | --- | --- | --- |
| Algorithm | P-value | Cliff's Delta | P-value | Cliff's Delta | P-value | Cliff's Delta |
| Apriori - FP-Growth | 0.031 | **1 (large)** | 0.031 | **1 (large)** | 0.031 | 1 (large) |
| Relim - FP-Growth | 0.079 | 0.068 (negligible) | 0.9 | 0.02 (negligible) | 0.1 | 0.06 (negligible) |
| Relim - Apriori | 0.031 | **-1 (large)** | 0.031 | **-1 (large)** | 0.031 | **-1 (large)** |
| Apriori - Eclat | 0.031 | **1 (large)** | 0.031 | **1 (large)** | 0.031 | **1 (large)** |
| Relim - Eclat | 0.1 | 0.1 (negligible) | 0.53 | 0.14 (negligible) | 0.06 | 0.13 (negligible) |
| FP-Growth - Eclat | 0.83 | 0.02 (negligible) | 0.31 | 0.13 (negligible) | 0.38 | 0.06 (negligible) |

Overall, we can conclude that Apriori outperforms Eclat, Relim, and FP-Growth in terms of Precision, Recall, and F-measure for the Kotlin project. Likewise, for execution time, Apriori and FP-Growth are the fastest algorithm while Relim and Eclat are about 15 and 27 times slower. We reject the null hypotheses H0-1, H0-2 and H0-3 for the comparisons of Apriori and FP-Growth, Relim and Apriori, and Apriori and Eclat.

Table 19: Average execution time for Kotlin.

| Algorithm | Time (seconds) |
|---|---|
| Apriori | 3,550 |
| FP-Growth | 1,030 |
| Relim | 15,918 |
| Eclat | 27,434 |

*6.6. Are there any differences between the four data mining techniques considered, i.e., Apriori, FP-Growth, Eclat and Relim in terms of their performance and execution time when recommending source code file changes for Rhino?*

Figure 19 shows that Eclat outperforms Relim and FP-Growth in terms of Precision.

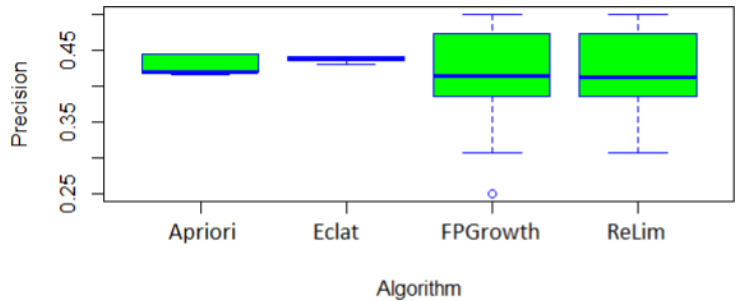

Figure 19: Boxplots of Precision for the Rhino project.

In terms of Recall, the boxplots in Figure 20 show that Relim and FP-Growth are slightly better than Eclat, while Eclat is slightly better than Apriori.

Finally, as illustrated in Figure 21, Eclat outperforms other data mining techniques.

As shown in Table 20, there are statistically significant differences between Relim - Eclat as well as FP-Growth and Eclat for the Rhino project (p-value =0.031 and Cliff's delta is small).

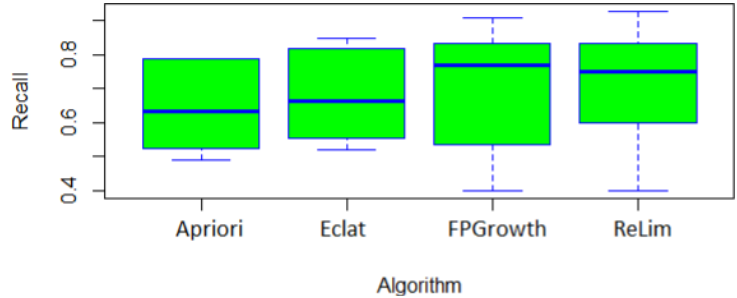

Figure 20: Boxplots of Recall for the Rhino project.

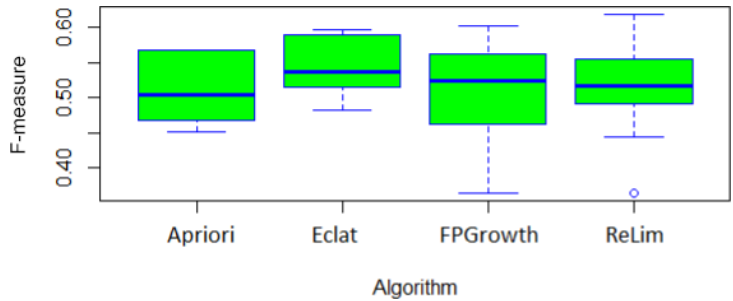

Figure 21: Boxplots of F-measure for the Rhino project.

Concerning Recall, Table 20 demonstrates that there are statistically significant differences for only one instance of comparison, namely Apriori and Eclat (p-value =0.031, while Cliff's delta is small).

Table 20 also shows a statistically significant difference in the F-measure for one instance only, namely Apriori and Eclat (p-value =0.031 with a medium Cliff's delta).

Considering the execution time, for Apriori, it only took 70 seconds to execute but all algorithms have close results (Table 21). Eclat ran in 132 seconds and FP-Growth and Relim have an execution time of 213 and 190 seconds.

In the case of Rhino, we conclude that there are statistically significant differences in some comparisons. Eclat, in particular, outperforms Relim and FP-Growth in terms of Precision (with a small effect size). Furthermore, it outperforms Apriori in terms of Recall (small effect size) and F-measure

Table 20: Adjusted P-values and Cliff's Delta for paired comparison between the four studied data mining algorithms for Rhino.

| Algorithm | Precision | | Recall | | F-measure | |
|---|---|---|---|---|---|---|
| | P-value | Cliff's Delta | P-value | Cliff's Delta | P-value | Cliff's Delta |
| Apriori - FP-Growth | 0.09 | 0.13888 (negligible) | 0.14 | -0.26 (small) | 0.84 | -0.000923 (negligible) |
| Relim - FP-Growth | 0.81 | 0.009259 (negligible) | 0.59 | -0.018 (negligible) | 0.9 | 0.046 (negligible) |
| Relim - Apriori | 0.06 | -0.13889 (negligible) | 0.14 | 0.26 (small) | 0.84 | 0.11 (negligible) |
| Apriori - Eclat | 0.16 | -0.33333 (medium) | 0.031 | -0.28 (small) | 0.031 | -0.44 (medium) |
| Relim - Eclat | 0.031 | -0.22222 (small) | 0.56 | 0.13 (negligible) | 0.56 | -0.17 (small) |
| FP-Growth - Eclat | 0.031 | -0.22222 (small) | 0.31 | 0.13 (negligible) | 0.69 | -0.22 (small) |

(medium effect size). Considering the execution time, While Apriori yields more efficiently, all algorithms have close execution time.

Table 21: Average execution time for Rhino.

| Algorithm | Time (seconds) |
|---|---|
| Apriori | 70 |
| FP-Growth | 213 |
| Relim | 190 |
| Eclat | 132 |

*6.7. Are there any differences between the four data mining techniques considered, i.e., Apriori, FP-Growth, Eclat and Relim in terms of their performance and execution time when recommending source code file changes for SWT?*

Figure 22 shows that Apriori outperforms Eclat, FP-Growth, and Relim.

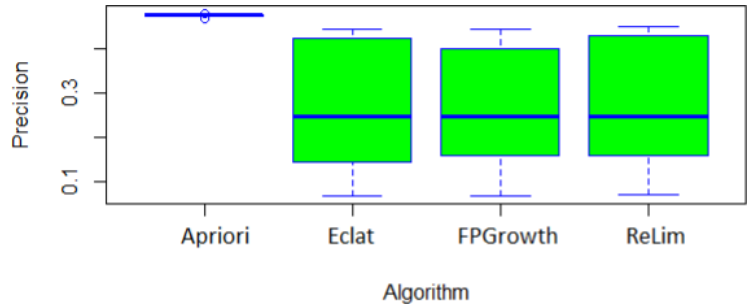

Figure 22: Boxplots of Precision for the SWT project.

Also, the boxplots in Figure 23 show that Apriori and Relim outperform Eclat and FP-Growth.

Figure 24 shows that Apriori outperforms other data mining techniques. Relim, on the other hand, outperforms FP-Growth and Eclat.

As shown in Table 22, there are statistically significant differences between Apriori and FP-Growth, Relim and Apriori, and Apriori and Eclat for the SWT project (p-value = 0.031 and Cliff's delta is large). There are also differences between Relim and FP-Growth (p-value = 0.014, but with a negligible effect size).

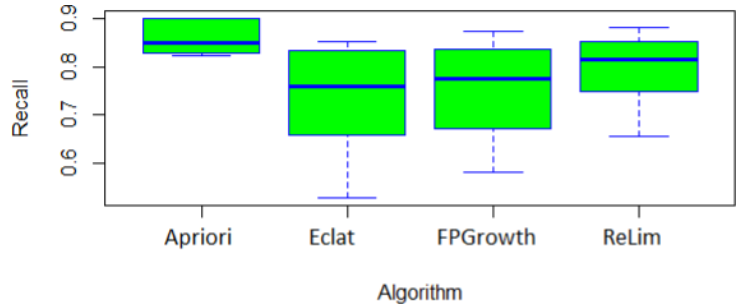

Figure 23: Boxplots of Recall for the SWT project.

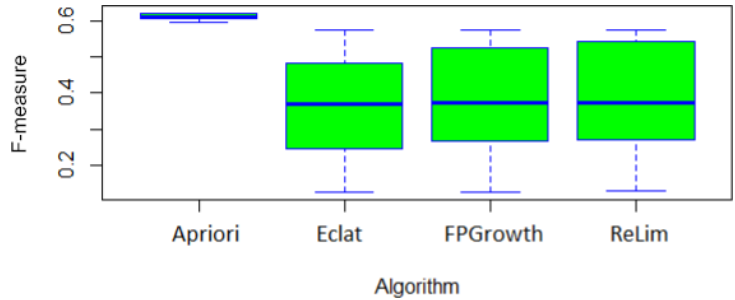

Figure 24: Boxplots of F-measure for the SWT project.

Table 22 shows that there are statistically significant differences in Recall between Relim and Eclat (p-value = 0.0045 and Cliff's delta is medium), as well as differences between Relim and FP-Growth (p-value = 0.01 and Cliff's delta is small).

In terms of the F-measure, we found statistically significant differences between Apriori and FP-Growth, Relim and Apriori, and Apriori and Eclat (p-value = 0.031, and Cliff's delta is large). Other comparisons, such as Relim and FP-Growth, as well as Relim and Eclat, produce statistical differences, but the effect size is not significant.

Talking about execution time, as shown in Table 23, Apriori is the best algorithm in terms of execution time with 329 second of execution time. while Eclat and Relim are the slowest ones with 6,374 and 5,250 seconds. Also it took 1,146 seconds for FP-Growth to generate the rules.

For SWT, we conclude that Apriori outperforms Eclat, Relim, and FP-Growth because significant differences with large effect-size for F-measure have been obtained. Furthermore, While Apriori's execution time is 329 seconds, Eclat is 20x slower than Apriori which makes it the most time-consuming algorithm in our research.

Table 22: Adjusted P-values and Cliff's Delta for paired comparison between the four studied data mining algorithms for SWT.

|  | Precision | | Recall | | F-measure | |
| --- | --- | --- | --- | --- | --- | --- |
| Algorithm | P-value | Cliff's Delta | P-value | Cliff's Delta | P-value | Cliff's Delta |
| Apriori - FP-Growth | 0.031 | **1 (large)** | 0.16 | **0.74 (large)** | 0.031 | **1 (large)** |
| Relim - FP-Growth | 0.014 | 0.1 (negligible) | 0.01 | 0.27 (small) | 0.012 | 0.1358025 (negligible) |
| Relim - Apriori | 0.031 | **-1 (large)** | 0.22 | **-0.51 (large)** | 0.031 | **-1 (large)** |
| Apriori - Eclat | 0.031 | **1 (large)** | 0.06 | **0.72 (large)** | 0.031 | **1 (large)** |
| Relim - Eclat | 0.12 | 0.07 (negligible) | 0.0045 | 0.35 (medium) | 0.018 | 0.1358025 (negligible) |
| FP-Growth - Eclat | 0.68 | -0.003 (negligible) | 0.13 | 0.099 (negligible) | 0.3 | 0.03703704 (negligible) |

Table 23: Average execution time for SWT.

| Algorithm | Time (seconds) |
|---|---|
| Apriori | 329 |
| FP-Growth | 1,146 |
| Relim | 5,250 |
| Eclat | 6,374 |

## 7. Discussion

To answer our main research question, for each project, we presented the calculated Precision, Recall and F-measure as well as execution time for all algorithms. Based on the boxplots and the information provided, the main findings can be summarised as follow.

Apriori yields better performances than Eclat, Relim and FP-Growth in particular in terms of precision and F-measure as the case for Kotlin, Elasticsearch and SWT projects, and sometimes in terms of recall too (e.g., Eclipse and Jabref projects).

Eclat shows better performances than Relim and FP-Growth in terms of precision and F-score for Rhino, it yields better performances than Apriori in terms of recall such as the case for Eclipse, as well as in terms of all performance measures as in the case of the Guava project. Additionally, Eclat performs better than FP-Growth for all performance measures in the Elasticsearch project. Relim and FP-Growth have shown better performances than Apriori in terms of recall for the Guava project.

Surprisingly, We observed an interesting pattern in these results. As previously mentioned in Table 7, Rhino is the smallest project in our dataset with only 3,383 records. After that, Guava has 5,654 records and finally, JabRef has 16,489 records. The other studied projects have more than 20,000 records, we refer to them here as large-scale projects. To better understand this pattern, we have combined the F-measure charts of three projects, Rhino, Guava and JabRef in Figure 25.

As illustrated in the figure, considering the median, for Rhino, Eclat yields better performance in terms of F-measure. For Guava, the results are much closer but still, Eclat and others are better than Apriori. Finally, for JabRef and all other projects Apriori outperforms other algorithms. We conclude that Apriori beats other algorithms when it comes to large projects with a huge amount of transactions. However, for smaller projects, recent algorithms specially Eclat outperform Apriori. This pattern is reported with F-measure as it is a harmonic mean of Precision and Recall but the same pattern is valid for them as well. Although in some cases the differences are not significant and may be considered negligible, we found this pattern interesting enough to report.

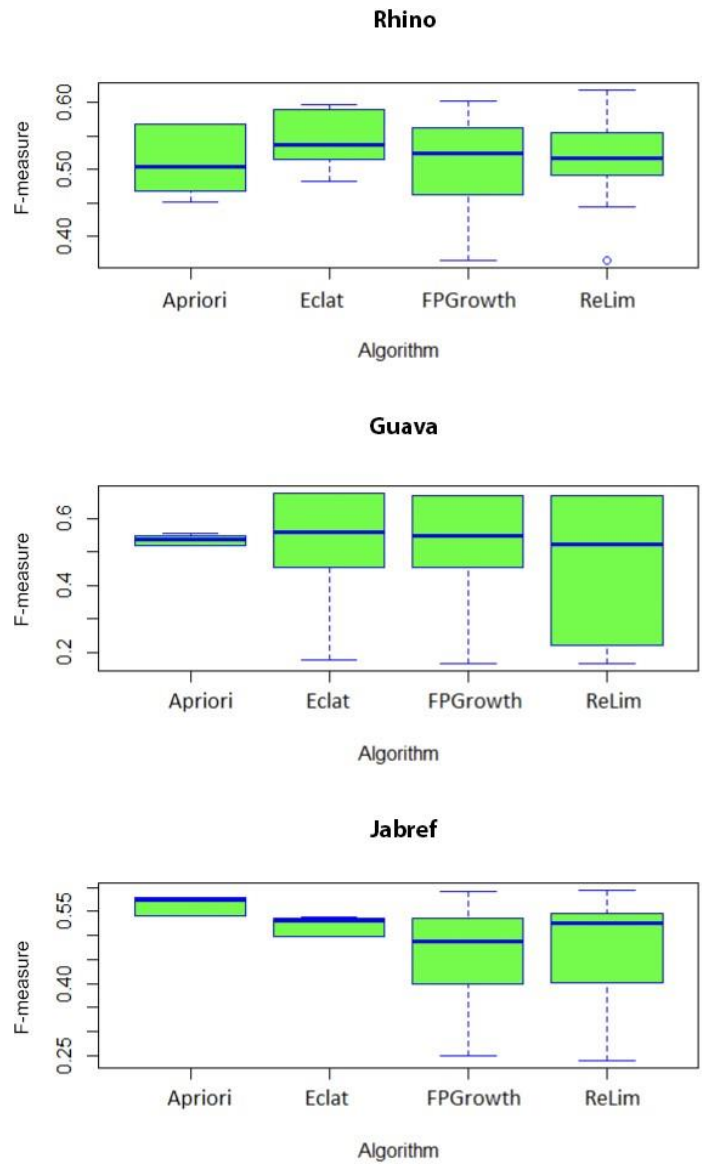

Figure 25: F-measure pattern of algorithms for different projects.

One possible explanation for the obtained results is that the Apriori algorithm generates fewer candidate sets of itemsets while running, resulting in better performance for projects such as Kotlin, Elasticsearch, and SWT, which have 15k, 12k, and 8k transactions, respectively, when compared to other projects with less than 3k transactions. Furthermore, Apriori uses optimized frequent itemsets generation strategies such as pruning techniques, which results in a reduction in the number of candidates and comparisons and, as previously stated, better performance for larger databases. However, when it comes to the smaller projects with fewer transactions, other data mining algorithms such as Eclat can compete with Apriori.

In terms of execution time, while generally, FP-Growth and Apriori are the fastest algorithm, Eclat has different behaviours. In small projects like Rhino, Eclat is close to other investigated algorithms. However, as the projects grow in terms of size, the gap between Eclat and other algorithms grows. This observation proves that Eclat is much more time-consuming compared to other algorithms for the high number of transactions, while for small datasets using Eclat may be beneficial. This is inline with the findings obtained by Chee et al. (2019) that we have summarized in section 2 (Table 4), which have pointed out to the fact that Eclat is an expensive algorithm in terms of memory space and processing time.

Considering the mentioned pattern and execution time for each algorithm, we can conclude that for small datasets using Eclat for recommending source code changes is beneficial. However, for large datasets with a high number of transactions, not only does Apriori has better performance but it is also much faster, *i.e.*, it took, on average, 2,037 seconds per 19,544 transactions. Moreover, with 1,386 seconds per 19,544 transactions, FP-Growth is the most efficient algorithm in terms of execution time especially for large-scale datasets.

We believe that the software engineering community and practitioners interested in applying data mining algorithms in their research can benefit from our findings in particular when choosing the proper data mining algorithm to use in the context of their software engineering tasks.

## 8. Threats to Validity

Although our study has been empirically and statistically evaluated, there are threats to its validity. The most common ones are covered in the following sections:

*8.1. Internal validity*

Internal validity is the degree to which a study's findings are reliable or accurate (Godwin et al., 2004). We provided empirical evidence of the differences between Frequent Pattern Mining (FPM) algorithms. Threats could be related to the data filtering we performed when filtering source code file changes. However, when dealing with such a process, we used widely-accepted filtering methods similar to those used in previous works. (Zimmermann et al., 2005; Ying et al., 2004).

*8.2. External validity*

External validity is the degree to which a study's findings are generalizable (Godwin et al., 2004). Eclipse, Elasticsearch, SWT, Kotlin, Guava, and JabRef are the seven real-world open-source projects examined in our research. The projects are not necessarily representative of all software projects. They differ in size and domains, however. It would be interesting to consider more projects in other contexts, such as the industrial field, in the future.

*8.3. Reliability validity*

Reliability is the degree to which a study produces comparable results under different conditions (Godwin et al., 2004). Eclipse, Elasticsearch, Rhino, SWT, Kotlin, Guava, and JabRef source code repositories are all open-source projects. The frame times of the code histories investigated by us are reported in the paper. Furthermore, we used the Python *Apyori* package to run Apriori and *pymining* package to run FP-Growth and Relim (with the parameters reported in the paper). In our online Appendix, we provide additional data that may be required to replicate our study[10].

---

[10]https://github.com/DataMining2022/JSS

*8.4. Conclusion validity*

The extent to which a study's conclusions are derived from sufficient data analysis and all research questions are accurately and logically answered is referred to as conclusion validity (García-Pérez, 2012). We conducted adequate tests for comparison to statistically reject the null hypotheses. We used non-parametric tests, which make no assumptions about the distributions of the data, and, in particular, the Wilcoxon pair-wise test. Furthermore, our conclusions are based not only on the presence of significant differences between pairs of data mining algorithms but also on the presence of the effect-size measure, i.e., the magnitude of the difference, between two algorithms. We also dealt with issues that arose when performing multiple Wilcoxon tests with the Holm correction.

## 9. Conclusion and Future Work

The use of data mining techniques for predicting patterns or recommending source code file changes is becoming more popular and has been quite common since the advent of big data.

Our focus in this paper is on four data mining algorithms that mine development history to recommend source code file changes. While some studies attempted to make recommendations for source code changes by utilizing simple data mining algorithms such as Apriori, they did not investigate other data mining techniques. Furthermore, they have not investigated various configurations of support and confidence, nor have they compared their work to any baseline or previous work. To the best of our knowledge, we are the first to propose recommending source code file changes by employing four distinct algorithms and comparing their performance across a wide range of configurations. We believe that such research can benefit both the research community and practitioners who want to use appropriate data mining algorithms in their research context.

In this paper, we used four advanced mining algorithms, *i.e.*, Apriori, FP-Growth, Eclat, and Relim on the development history of seven widely used open-source projects, *i.e.*, Eclipse, Elasticsearch, Rhino, SWT, Kotlin, Guava, and JabRef. We compared the performance and results of the four investigated algorithms in terms of Precision, Recall, and F-measure and their execution time.

Our findings show that Apriori outperforms Eclat, Relim, and FP-Growth in terms of Precision and F-measure, as is the case for Kotlin, Elasticsearch, and SWT projects, and sometimes in terms of Recall as well, *e.g.*, Eclipse and JabRef projects. Eclat outperforms Apriori in terms of Recall, as seen in the case of Eclipse, as well as overall performance, as seen in the Guava project. Additionally, Eclat outperforms FP-Growth across all performance metrics in the Elasticsearch project. Relim and FP-Growth, on the other hand, outperformed Apriori in terms of Recall for the Guava project. When it comes to execution time, While Apriori is the fastest algorithm in general, Eclat has unique characteristics. In small-scale projects such as Rhino, Eclat is comparable to other algorithms. However, as projects grow in size, the gap between Eclat and others grows.

Considering the aforementioned points, We can conclude that Apriori has better performance for large databases with a large amount of data. However, for smaller projects with less number of transactions, other data mining algorithms specially Eclat can be promising.

For future works, we plan to investigate more projects with programming languages other than Java. We also intend to improve the quality of our transactions by applying various filters to our development history and taking into account metrics on the quality of our transactions to identify only those that are relevant to the developer's task at hand. Finally, one possibility is to create new models of source code changes that take into account the temporal aspect of change history, specifically the order of the source code changes. Several sequential pattern mining algorithms have already been used in areas such as biology to study DNA sequences and could be leveraged as heuristics in addition to the approaches considered in this paper.

## Acknowledgements

Baysal would like to acknowledge the support of the Natural Sciences and Engineering Research Council of Canada (NSERC), RGPIN-2021-03809.

Guerrouj would like to acknowledge the support of the Natural Sciences and Engineering Research Council of Canada (NSERC), RGPIN-4712-2016.